\documentclass[aps,prd,
preprint,groupedaddress,superscriptaddress]{revtex4}

\usepackage{graphicx}
\usepackage{amsmath}
\usepackage{bm}
\usepackage{color}
\usepackage{amssymb,enumerate}

\usepackage[bookmarks=true]{hyperref}
\usepackage[bulletsep]{collect}

\def\ee{\end{eqnarray}}

\def\=:{=\hspace{-.7em}\raisebox{1.1ex}{.}\hspace{.1em}\raisebox{-0.2ex}{.} }

\def\ee{\end{eqnarray}}

\def\=:{=\hspace{-.7em}\raisebox{1.1ex}{.}\hspace{.1em}\raisebox{-0.2ex}{.} }

\newcommand {\beq}{\begin{eqnarray}}
\newcommand {\eeq}{\end{eqnarray}}
\newcommand {\non}{\nonumber\\}

\newcommand {\1}[1]{\frac{1}{#1}}

\newcommand {\del}{\partial}

\begin{document}


\title{Quantum Exact Non-Abelian Vortices in 
Non-relativistic Theories
}


\author{Muneto Nitta}
\affiliation{
Department of Physics, and Research and Education Center for Natural 
Sciences, Keio University, Hiyoshi 4-1-1, Yokohama, Kanagawa 223-8521, Japan
}
\author{Shun Uchino}
\affiliation{DPMC-MaNEP, University of Geneva, 24 Quai Ernest-Ansermet, 
CH-1211 Geneva, Switzerland
}
\author{Walter Vinci}
\affiliation{London Centre for Nanotechnology and Computer Science, 
University College London, 17-19 Gordon Street,
London,  WC1H 0AH, United Kingdom 
}


\date{\today}
\begin{abstract}
Non-Abelian vortices arise when a non-Abelian global symmetry is exact in the ground state but spontaneously 
broken in the vicinity of their cores. 
In this case, there appear (non-Abelian)  Nambu-Goldstone (NG) modes 
confined and propagating along the vortex. In relativistic theories, the Coleman-Mermin-Wagner theorem 
forbids the existence of 
a spontaneous symmetry breaking, or a long-range order, 
in 1+1 dimensions: quantum corrections restore the symmetry along the vortex and the NG modes acquire a mass gap. We show that in non-relativistic theories NG modes 
with quadratic dispersion relation confined on a vortex can remain gapless at quantum level.  
We provide a concrete and experimentally realizable example of a three-component 
Bose-Einstein condensate with 
$U(1) \times U(2)$ symmetry. 
We first show, at the classical level, 
the existence of $S^3 \simeq S^1 \ltimes S^2$  ($S^1$ fibered over $S^2$)   
NG modes associated to the breaking 
$U(2) \rightarrow  U(1)$ on vortices, 
where $S^1$ and $S^2$ correspond to  type I and II NG modes, respectively. 
We then show, by using a Bethe ansatz technique, that
the $U(1)$ symmetry is restored, while the $SU(2)$ symmery remains broken non-pertubatively at quantum level. 
Accordingly, the $U(1)$ NG mode turns into 
a $c=1$ conformal field theory,  the Tomonaga-Luttinger liquid, 
while the $S^2$ NG mode remains gapless,  
describing a  ferromagnetic liquid. 
This  allows the vortex to be 
genuinely non-Abelian at quantum level.

\end{abstract}

\maketitle
\section{Introduction}

Vortices play a prominent role in various areas of physics such as  superfluidity, superconductivity,
magnetic materials, nematic liquids, fractional quantum Hall effect, topological quantum computation, quantum field theory and cosmology. 
In particular in high energy physics, 
much attention has been recently paid to 
 non-Abelian vortices.   
Such objects arise when a non-Abelian symmetry is spontaneously broken in the vicinity of the vortex. The  
gapless Nambu-Goldstone (NG) modes associated to this symmetry breaking are then confined on the vortex, where they can propagate.
%
%
%
%
Non-Abelian vortices were first 
discovered in supersymmetric gauge theories 
\cite{Hanany:2003hp,Auzzi:2003fs,Eto:2005yh,Eto:2006cx,Eto:2006pg} 
(see Refs.~\cite{Eto:2006pg,Shifman:2007ce,
2009supersymmetric,Tong:2008qd} for a review), 
where their presence elegantly explains a long standing problem
about quantum physics in different dimensions, {\it i.e.} 
similarities in the non-perturbative dynamics of 
sigma models in 1+1 dimensions 
and Yang-Mills theory in 3+1 dimensions ( {\it e.g.}
existence of dynamical mass gap, 
asymptotic freedom,  instantons). 
Non-Abelian vortices also appear 
in high density quark matter (in this context also known as color flux tubes)
\cite{Balachandran:2005ev,Nakano:2007dr,Eto:2009kg,Eto:2009bh,Eto:2009tr} (see Ref.~\cite{Eto:2013hoa} for a review). 
In this case, a Coleman-Weinberg quantum potential 
arising on the vortex world-sheet leads to the presence of monopole mesons confined by  color flux tubes 
\cite{Gorsky:2011hd,Eto:2011mk}.
This phenomenon explains a quark-hadron duality 
between low and high density QCD, {\it i.e.}  
quarks are confined when monopoles are condensed 
at low density while 
monopoles are confined when quarks are condensed 
at high density 
\cite{Eto:2011mk}. 
The examples mentioned above correspond to relativistic systems. 
In such cases NG modes acquire mass gaps 
through quantum corrections \cite{Shifman:2004dr,Hanany:2004ea}    consistently  with 
the Coleman-Mermin-Wagner theorem \cite{Coleman:1973ci,PhysRevLett.17.1133} 
which forbids the existence of 
a spontaneous symmetry breaking or a long-range order 
in 1+1 dimensions. 
In this sense, non-Abelian vortices studied thus far in relativistic theories 
are Abelianized at quantum level \cite{Dorey:1999zk,Shifman:2004dr}. 

On the other hand, 
 the situation can be 
different in non-relativistic theories.
There are two kinds of NG modes, 
 type-I and II, with linear and quadratic dispersions  
respectively \cite{Nielsen:1975hm}. 
This distinction is deeply connected with the geometry of the space parametrized by the NG modes 
\cite{Watanabe:2012hr,Hidaka:2012ym}; 
one type-II NG mode corresponds 
to {\it two} broken generators, 
while one type-I NG mode corresponds to 
one broken generator. In the relativistic case only type-I NG are allowed.

In this paper, 
we show that type-II NG modes confined in the core of a vortex 
remain gapless at quantum level, 
providing a {\it quantum exact} non-Abelian vortex.
We consider  a concrete example of a three-component 
Bose-Einstein condensate (BEC) with $U(2) \times U(1)$ symmetry, 
in which there is partial phase separation between one (immiscible) component and the other two (miscible) components.  
Our setup can be realized in experiments of 
ultracold atomic gases;
(Almost) $U(2)$ symmetric two-component BECs 
were already realized by 
ultracold 
$^{87}$Rb \cite{PhysRevLett.81.1539} and 
$^{23}$Na \cite{PhysRevLett.82.2228} atoms. 
If one mixes additional 
atom with repulsive interaction with these atoms, 
one can experimentally realize our set-up.
In this system, 
we find that there appear 
an $S^1$ type-I NG mode and 
an $S^2 \simeq {\mathbb C}P^1$ type-II NG mode 
in a vortex core. 
By noting that this vortex acts as a 1+1 dimensional trap 
for the two miscible components, we describe the modes living in the vortex core as a 1+1 dimensional two-component Bose gas with $U(2)$ symmetry. By studying the exact solution of this system through a Bethe ansatz technique, we show that at quantum level  
the $U(1)$ symmetry recovers and the $U(1)$ sector can be described by
a conformal field theory (CFT) with the conformal charge $c=1$ or 
a Tomonaga-Luttinger liquid \cite{giamarchi2003quantum,Essler.etal/book.2010}. On the other hand, 
 the $SU(2)$ symmetry 
breaking holds at quantum level 
and the $S^2$ NG mode remains gapless, 
describing a ferromagnetic liquid \cite{RevModPhys.83.1405}.

\section{Effective Lagrangian approach}

We consider a three-component BEC with $U(1)\times U(2)$ symmetry. 
The Lagrangian for the Gross-Pitaevskii equation is
\begin{eqnarray}
&& \hspace{-0.5cm}
 \mathcal L =  i \hbar \psi_0^\dagger \dot\psi_0+ i \hbar \Psi^\dagger \dot\Psi-\frac{\hbar^2}{2 m_0} \nabla_i \psi_0^\dagger \nabla_i \psi_0 -\frac{\hbar^2}{2 m} \nabla_i \Psi^\dagger \nabla_i \Psi  \nonumber \\
&&+  \mu_0 |\psi_0|^2 +  \mu |\Psi|^2 - \frac12 \lambda_0 |\psi_0|^4- \frac12 \lambda |\Psi|^4 -  \kappa  |\psi_0|^2 |\Psi|^2\,, \non
\end{eqnarray}
which is written in terms of the  condensate wave functions
$(\psi_0,\Psi^T)$, where $\Psi$ is a two-component condensate. 
Here, $\mu$ and $\mu_0$ are chemical potentials, 
and $\lambda$ ($\lambda_0$) and 
$\kappa$ are intracomponent and intercomponent couplings, 
respectively. 
The system above has a stable minimum when $\lambda_0 \lambda -\kappa^2>0$. We consider the phase separation between $\psi_0$ (immiscible component) and 
$\Psi$ (miscible components), with a non-vanishing expectation value for $\psi_0$, which is obtained when $ \mu_0 \lambda -\mu \kappa>0$, $\mu \lambda_0 -\mu_0 \kappa < 0$ and $ \mu^2 \lambda_0 -\mu_0^2 \lambda <0$. In this situation, the ground state is  $(\psi_0,\Psi^T) = (v,0,0)$ 
with $v^2= \mu_0/\lambda_0$, 
where the symmetry is spontaneously broken down to 
$U(2)$. 

\begin{figure}
\centering
\includegraphics[width=0.9\columnwidth]{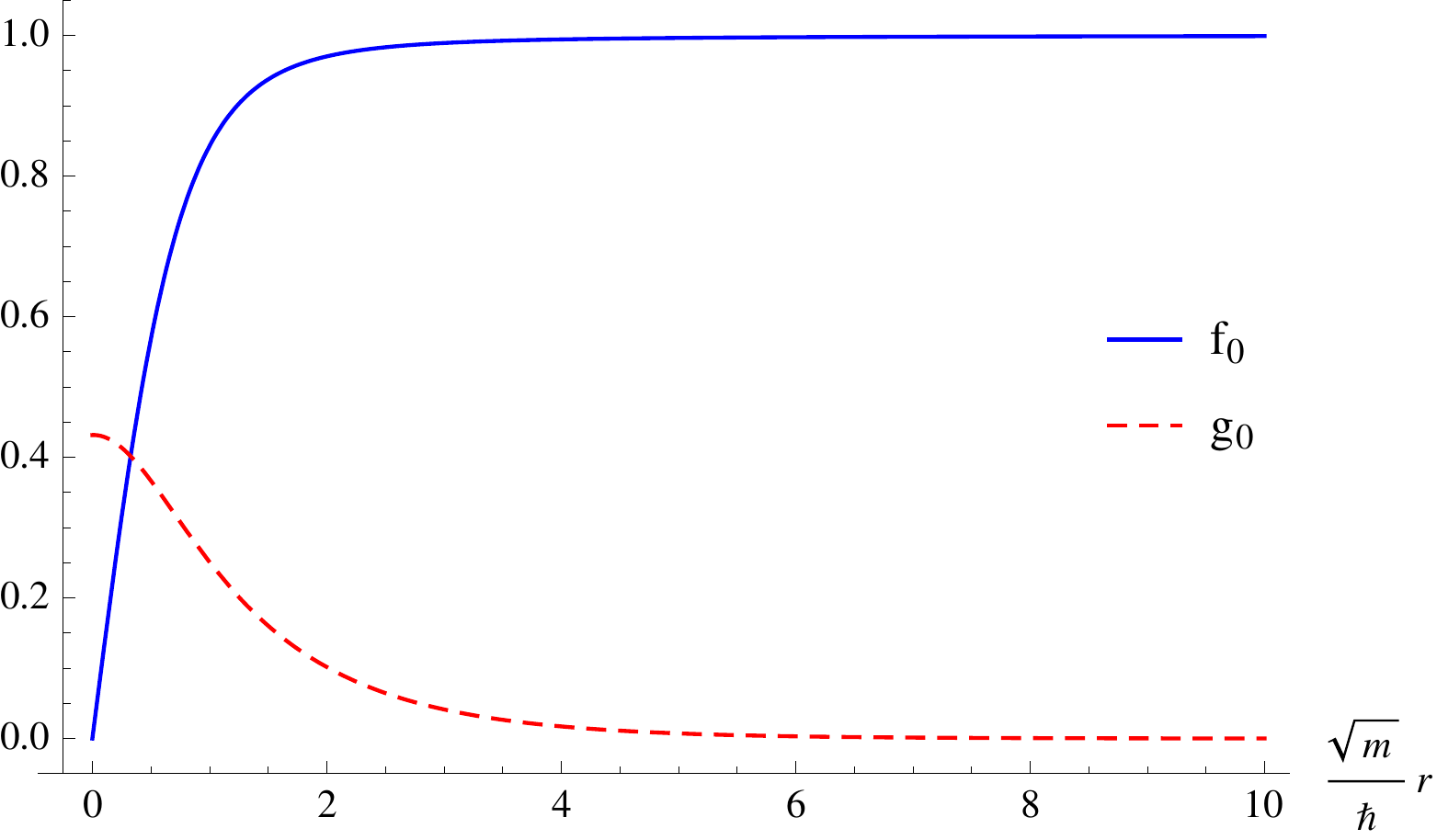} 
\caption{Numerical solution for the profile functions $f_0$ and $g_0$. $\mu=2$, $\mu_0 = \lambda = \lambda_0=3$, $\kappa = 2.3$, $m = m_0$.}
\label{fig:profiles}
\end{figure}

In the cylindrical coordinates $(r,\theta,z)$, 
a vortex along the $z$-axis is given by 
\begin{equation}
\left(
\begin{array}{c}
  \psi_0    \\
   \Psi
\end{array}
\right)
= \left(
\begin{array}{ccc}
 f_0(r,\theta)e^{i\theta}   \\
   g_0(r,\theta) \eta
\end{array}
\right),\quad \eta^\dagger \eta =1, 
\end{equation}
with the boundary conditions 
$f_0 \to v$ and $g_0 \to 0$ at $r \to \infty$ 
and 
$f_0 \to 0$ and $g_0' \to 0$ at $r \to 0$, see Fig.~\ref{fig:profiles}. 
Here,  $\eta$ is a two-component complex constant that parameterizes $S^3$, and 
identify the NG modes 
associated with the ground state symmetry breaking
$U(2) \rightarrow U(1)$ that takes place 
in the vicinity of the vortex core:
$U(2)/U(1) \simeq SU(2) \simeq S^3$.

We construct a low-energy effective theory 
of the vortex through the moduli approximation
\cite{Manton:1981mp,Eto:2006uw}
introducing  a  world-sheet coordinate  ($t,z$)  dependence
of the vortex configuration  
\begin{equation}
\left(
\begin{array}{c}
  \psi_0    \\
   \Psi
\end{array}
\right)
= \left(
\begin{array}{ccc}
 f(r,\theta,z,t)e^{i\theta}   \\
   g(r,\theta,z,t) \eta(z,t)
\end{array}
\right),\quad \eta^\dagger \eta =1 .\label{eq:ansatz}
\end{equation}
Notice that we have included a world-sheet coordinate dependence on the profile functions too. This is equivalent to the inclusion of  ``Higgs" excitations $H$ and $h$ in the calculation:
\beq
&& f(r,\theta,z,t) \equiv f_0(r) + H(r,\theta,z,t), \quad \non
&& g(r,\theta,z,t) \equiv g_0(r) + h(r,\theta,z,t) .
\eeq
This modification to the standard approach in deriving effective theories in the relativistic case is crucial. The field $H$ is a massive bulk mode, while the field $h$ is a lighter non-propagating mode that contributes to the effective action for the zero-modes $\eta$. 

 By substituting Eq.~(\ref{eq:ansatz}) in the Lagrangian we get:
 \begin{eqnarray}
\mathcal L & = &i\hbar g^2 \eta^\dagger \dot \eta \nonumber \\
&-&\frac{\hbar^2}{2 m_0} \left[  f'^2+\frac{1}{r^2}f^2+ \frac{1}{r^2}(\nabla_\theta f)^2 + (\nabla_z f)^2 \right]\nonumber \\
& -& \frac{\hbar^2}{2 m} \left[  g'^2+ \frac{1}{r^2}(\nabla_\theta g)^2 + (\nabla_z g)^2 +g^2 |\nabla_z \eta|^2\right] \nonumber \\
&+&  \mu_0 f^2 +  \mu g^2 - \frac12 \lambda_0 f^4- \frac12 \lambda g^4 -  \kappa  f^2 g^2,
\label{eq:lagrangian}
 \end{eqnarray} 
from whose variation one obtains the equations of motion for $f$ and $g$:
\begin{eqnarray}
 &&0= -\frac{\hbar^2}{2m_0}\left[ \frac{1}{r^2}f -\triangle f   \right] +  \mu_0 f  -  \lambda_0 f^3  -  \kappa  f g^2  \nonumber \\
 &&0=
 i\hbar   g \eta^\dagger \dot \eta -\frac{\hbar^2}{2m}\left[ g |\nabla_z \eta|^2-\triangle g \right] 
+  \mu g  -  \lambda g^3  -  \kappa  g f^2 .\non
\label{eq:equations}
\end{eqnarray}
In the derivation of the effective action for the moduli $\eta$ we will keep only terms up to  second derivatives in the world-sheet coordinates. This allows us to consider an expansion of the equations above in powers of these coordinates.
An important observation is that, thanks to the (static) equation of motions, the corrections to the effective Lagrangian in Eq.~(\ref{eq:lagrangian}) enter quadratically at least. This allows us to calculate all terms in the effective action, up the the second order in the world-sheet coordinates, by just calculating corrections to the profile functions up to the first order. As can be seen by inspecting Eq.~(\ref{eq:equations}), first order corrections will only involve time derivatives of the moduli $\eta$. We proceed using the following expansion:
\begin{eqnarray}
 &&g = g_0+ i\hbar    \eta^\dagger \dot \eta \, g_1+\mathcal O(\partial_t^2,\partial_z^2), \nonumber  \\
  &&f = f_0+  i\hbar    \eta^\dagger \dot \eta \,f_1+\mathcal O(\partial_t^2,\partial_z^2). 
\end{eqnarray}
We first consider the zeroth order, corresponding to the static case, and study the vortex profile functions $f_0$ and $g_0$. The equations for these two functions correspond to those in Eq.~(\ref{eq:equations}) without the world-sheet derivatives. A numerical example  can be  seen in Fig.~\ref{fig:profiles}. We can see how the vortex built in the immiscible  component acts as a one dimensional trap for the  two miscible components. This is the first crucial observation of our paper. The asymptotic behavior of the profile functions can be found analytically by studying the  expansion $f_0 = v-\delta f$, $g_0 =\delta g$ of Eq.~(\ref{eq:equations}) from which we get:
\begin{eqnarray}
 f_0 &\sim &v-\frac{1}{2\sqrt{\mu_0 \lambda_0}}\frac{1}{r^2},  \nonumber \\
 g_0 &\sim& \frac{c_1}{\sqrt r} e^{-  \gamma r}, \quad   \gamma \equiv \sqrt{\frac{-\lambda_0 \mu +\kappa \mu_0 }{\lambda_0}}\,.
 \label{eq:asimptotic}
\end{eqnarray}
As common in systems with only global symmetries, the profile function $f_0$ for the condensate supporting the winding decays polynomially. On the other hand the expression above shows the profile function for the trapped components $g_0$ decays exponentially. This is our second important observation. The exponential decay allows us to search for normalizable zero-modes propagating on the vortex line, 
so that 
the 1+1 effective action is finite on an infinite volume.

We proceed now to the calculation of the effective action with the derivation of the higher order corrections $ f_{1}$ and $ g_{1}$. The  expansion of Eq.~(\ref{eq:equations}) in powers of the world-sheet coordinates gives new differential linear equations for the variables  $ f_{1}$ and $ g_{1}$ 
(see Appendix \ref{app:eff-ac}).  
Since these equations are linear in the first order corrections, $f_1(r)$ end $g_1(r)$ depend only on the zeroth order profile functions $f_0$ and $g_0$. Their numerical evaluation can be found in Fig.~\ref{fig:firstorderprofiles} in Appendix \ref{app:eff-ac}. Substituting the expressions $g = g_0+ i\hbar    \eta^\dagger \dot \eta \, g_1$ and $f = f_0+  i\hbar    \eta^\dagger \dot \eta \,f_1$ in the original Lagrangian (\ref{eq:lagrangian}), and remembering to keep track of the second order terms only, we obtain:
\begin{eqnarray}
\mathcal L & = &i\hbar g_0^2 \eta^\dagger \dot \eta  + K\,  \hbar^2  (\eta^\dagger \dot \eta)^2  -\frac{\hbar^2}{2m} g_0^2 |\nabla_z \eta|^2 \nonumber \\
K &\equiv &-   \big(  \mu_0 f_1^2+ \mu g_1^2  -3 \lambda_0 f_0 f_1^2-3 \lambda g_0 g_1^2 \non 
&  &-  4\kappa f_0 g_0 f_1g_1-\kappa  g_0^2 f_1^2-\kappa  f_0^2 g_1^2 \big) .
 \end{eqnarray} 
The effective action is obtained after integrating over the transverse direction $r$, this can be easily done numerically with the knowledge of the profile functions $f_{0,1}$ and $g_{0,1}$. The whole calculation has assumed only a derivative expansion  in powers of the world-sheet derivatives, keeping terms up to the second order:
\begin{eqnarray}
&& \mathcal L_{\rm eff} =   \alpha  i \hbar \eta^\dagger \dot \eta -  \alpha \frac{\hbar^2}{2m}  |\nabla_z \eta|^2 - \beta  \hbar^2  (\eta^\dagger \dot \eta)^2\nonumber \\
&& \alpha  =  \int_0^\infty 2 \pi r dr g_0^2 , \quad  \beta  =  \int_0^\infty 2 \pi r dr  K \,.
\label{eq:coeffs}
\end{eqnarray}
 Since the function $g_0$ is exponentially decaying, the coefficient $\alpha$ is finite. Moreover, notice that the first order correction $i \hbar g_0 \eta^\dagger \dot \eta$ is exponentially small at large distances, which implies that also the  functions  $f_1$ and $g_1$ decay exponentially. As a consequence, the coefficient $\beta$ is  finite and of the same order of $\alpha$. 
This means that the effective action is finite and the zero modes are truly confined on the vortex. The coefficients $\alpha$ and $\beta$ have been calculated numerically in  
Appendix \ref{app:eff-ac} 
for the special choice of parameters shown in  Fig.~\ref{fig:profiles}.

In order to see the dispersions of the NG modes, 
we recall that the geometry of the NG modes is 
$S^3 \simeq S^1 \ltimes S^2$, 
where $S^1$ is fibered over $S^2 \simeq {\mathbb C}P^1$.
To make this structure explicit, 
we take a parameterization 
\beq
 \eta = e^{i \varphi} n
\eeq
with $n^\dagger n=1$, 
where $\varphi$ and $n$ parametrize 
$S^1$ and $S^2$. 
Then, the effective Lagrangian can be rewritten 
as  
\beq
&& {\cal L}_{\rm eff.} 
=
- \beta \hbar^2 \left(\dot \varphi  - {i\over 2} (n^\dagger \dot n - \dot n^\dagger n)\right)^2 \non
&&
- \alpha {\hbar^2 \over 2m} 
\left( \del_z \varphi 
 - {i\over 2} (n^\dagger \del_z n - \del_z n^\dagger n)  \right)^2
\non
&&
 + 2i \alpha \hbar (n^\dagger \dot n - \dot n^\dagger n) 
 - \alpha {\hbar^2 \over 2m} 
 \left((\del_z n)^2  - (n^\dagger \del_z n)^2 \right), 
\eeq
which shows that 
$\varphi$ and $n$ are type-I and II NG modes 
with linear and quadratic dispersions, respectively. 
Here, remarkably 
$n$ describes one type-II NG mode corresponding to 
the two broken generators of $S^2$.
In addition, there appears 
one type-II translational zero mode 
corresponding to two broken translational symmetries  
\cite{Kobayashi:2013gba}.

\section{Quantum exact gapless modes}

 In the effective theory approach, we have obtained 
gapless modes at classical level. 
In order to discuss whether these modes
 remain gapless quantum mechanically, 
we should take into account massive modes 
and quartic interaction. 
As we have noticed, the vortex  in 
the first component 
can be seen as a trap 
for the second and third components, 
resulting in 
two-component Bose gases with $U(2)$ symmetry in 1+1 dimension. 
Here, we study such two-component Bose gases at quantum level.

Two-component Bose gases with $U(2)$ symmetry in 1+1 dimension 
are also known as Yang-Gaudin model 
\cite{gaudin1967systeme,PhysRevLett.19.1312}.
Including a nonlinear interaction, the Hamiltonian  in
units of $\hbar^2/(2m)$
is given by
\beq
H=\int_{0}^{L} dz \left(
\sum_{\sigma=1,2}\partial_z \Psi^{\dagger}_{\sigma}\partial_z \Psi_{\sigma}
- \mu \rho
+\lambda\rho^2\right),
\eeq
where $\rho=\sum_{\sigma}\Psi^{\dagger}_{\sigma}\Psi_{\sigma}$
is the density,
and $L$ is a system size along the $z$-direction. 
The effective parameters $\lambda$ and $\mu$ 
can be related to the original parameters through 
a renormalization group, but here we do not need 
explicit relations. 
Since there is no spin-dependent coupling, the Hamiltonian has
$U(2)$ symmetry in which we can use the Bethe ansatz technique
\cite{deguchi2000thermodynamics,Essler.etal/book.2010}.
The resultant Bethe ansatz equations 
are \cite{li2003exact,
PhysRevLett.95.150402} (See Appendix \ref{App:Bethe})
\beq
k_{j}L=2\pi I_{j}-\sum_{l=1}^{N}\Theta\left(\frac{k_j-k_l}{\lambda}\right)
+\sum_{\beta=1}^M\Theta\left(\frac{2k_j-2l_{\beta}}{\lambda}\right)
\label{eq:bae1}
\\
2\pi
J_{\alpha}=\sum_{j=1}^{N}\Theta\left(\frac{2l_{\alpha}-2k_j}{\lambda}
\right)-\sum_{\beta=1}^M\Theta\left(\frac{l_{\alpha}
-l_{\beta}}{\lambda}\right)
\label{eq:bae2}
\eeq
where $\Theta(x)=2\tan^{-1}(x)$, and
$N$ and $M$ represent the total number of the particles 
and particle number for the component ``2'', respectively.
$k_{j}$ and $l_{\alpha}$ are called quasimomenta and isospin
rapidities originating from $SU(2)$ symmetry in the system.
$I_{j}$ and $J_{\alpha}$ take integer or half-integer values,
depending on whether $N-M$ is odd or even.
Then, the total energy and momentum are given by
\beq
E=\sum_{j=i}^{N}(k^2_{j}-\mu),\quad
P=\sum_{j=1}^Nk_{j}\,.
\label{eq:momentum}
\eeq
By using $I_j$ and $J_{\alpha}$, 
the ground state can be characterized as \cite{li2003exact}
\beq
\{I_{j}\}=\{-(N-1)/2,\cdots,(N-1)/2\}, \quad M=0.
\eeq
In the thermodynamic limit, the above Bethe ansatz equations are then
 reduced to
\beq
\rho(k)=\frac{1}{2\pi}+\frac{1}{\pi}\int_{-\Lambda}^{\Lambda}dk'
\frac{\lambda\rho(k')}{\lambda^2+(k-k')^2}
\label{eq:lieb-liniger},
\eeq
where $\rho(k)$ is the density of quasimomenta, 
and cut-off $\Lambda$ is determined by the density of the system
$N/L=\int_{-\Lambda}^{\Lambda}\rho(k)$. 
This equation corresponds
to that in Lieb-Liniger gas \cite{PhysRev.130.1605,PhysRev.130.1616}.
Thus, as far as  $S^1$ sector is concerned, 
the low-energy excitation is linear gapless characterized by 
the $c=1$ CFT while there is no BEC.
In addition, $M=0$ means that the ground state is ferromagnetic,
which is consistent with a general theorem that
a ground state of spinful bosons without spin-dependent
interaction is always polarized \cite{PhysRevLett.89.220403} .

We next consider the nature of the isospin excitation originating from
$S^2$ manifold.
Although  analytic results can be obtained for weak and strong coupling limits,
we need to numerically determine it
for general couplings.
To this end, we consider
 a dressed energy formalism in the Bethe ansatz \cite{Essler.etal/book.2010}.
In our model, we can obtain
\beq
\epsilon_{c}(k)=k^2-\mu +\frac{1}{\pi}\int_{-\Lambda}^{\Lambda}
dk'\frac{\lambda\epsilon_{c}(k')}{\lambda^2+(k-k')^2},
\label{eq:ds-charge}\\
\epsilon_{s}(l)=-\frac{1}{2\pi}\int_{-\Lambda}^{\Lambda}
dk'\frac{\lambda\epsilon_{c}(k')}{(\lambda/2)^2+(l-k')^2},
\label{eq:ds-spin}
\eeq 
where $\mu$ is the chemical potential and we introduced
the dressed energy $\epsilon_c(k)$ and $\epsilon_s(l)$,
each of which represents $S^1$ and $S^2$ excitations.
The derivation of the above equations is given in 
Appendix \ref{App:Bethe}.
In the low-energy isospin excitation, an isospin rapidity $l$  
is related to the \textit{real} momentum of the isospin 
excitation $p$ as follows:
\beq
p=\int_{-\Lambda}^{\Lambda}dk'\rho(k')\left[
\Theta\left(\frac{2k'-2l}{\lambda}\right)+
\pi\right].
\label{eq:spin-momentum}
\eeq
By solving a couple of the above equations,
we can confirm that
the isospin excitation is gapless quadratic one \cite{li2003exact,
PhysRevLett.95.150402}.
This indicates that 
a type-II NG mode, called a magnon, 
emerges for the isospin sector
due to a spontaneous symmetry breaking of $SU(2)$.

\section{Summary and Discussion}

We have shown that a non-Abelian vortex 
remains non-Abelian at the quantum level 
when trapped NG modes are of type-II, \emph{e.g.} the low-energy behavior
of the NG modes  is robust against quantum fluctuations. 
We have worked out an explicit example of  
a three component BEC 
with $U(2) \times U(1)$ symmetry. 
At the classical level,  there appear in the 1+1 dimensional 
low-energy effective theory of the vortex
both type I and II NG modes. $\varphi$ and $n$,
with linear and quadratic dispersions respectively,  are associated to the fiber $S^1$ and the base $S^2$ of the moduli space $S^3 \simeq S^1 \ltimes S^2$ , where  $S^3$ is associated with the symmetry breaking induced by a vortex solution
$U(2) \to U(1)$.    At quantum level,  however,
the $U(1)$ symmetry is recovered
while $\varphi$ remains gapless with a linear dispersion. On the other hand, the symmetry 
breaking  $SU(2) \to U(1)$  holds at quantum level 
and $n$ remains gapless with a quadratic dispersion.
We then should interpret the type-I NG mode $\varphi$
as the $c=1$ CFT mode at quantum level
since there is no BEC,  that is, a Tomonaga-Luttinger liquid.
At the same time, the type-II NG mode  $n$ remains at quantum level and is interpreted as a ferromagnetic liquid.

An intuitive understanding of the above is as follows.
As far as the low-energy physics is concerned where
$S^1$ and isospin sectors are expected to be separated,
the kinetic Hamiltonians in $S^1$ and isospin sectors
are Lorentz and Galilei invariant, 
respectively, thanks to their dispersion rules. 
Then, while the spontaneous symmetry breaking in the $S^1$ sector
is forbidden by the Coleman-Mermin-Wagner theorem,
there still exists a linear gapless mode since 
the fixed point of the system is the so-called Tomogana-Luttinger liquid
where all of the interactions that can yield gaps in the system are irrelevant.
On the other hand, the spin sector
is purely non-relativistic in the sense that Lorentz
symmetry does not emerge even in the low-energy physics. Since one of the conditions of the Coleman-Mermin-Wagner theorem is Lorentz invariance, spontaneous symmetry breaking in the isospin sector is not forbidden and the type-II NG mode remains at the quantum level as prescribed by the NG theorem.

We can also consider the  case where the $U(2)$  symmetry is explicitly broken to $U(1)^2$, but the trapped components are still miscible. This can be obtained with an additional term of the type $-{\cal L}_{\rm int} = \lambda_{12} |\Psi_1|^2|\Psi_2|^2+
\1{2} \lambda_{11}|\Psi_1|^4 + \1{2} \lambda_{22}|\Psi_2|^4$, where $\lambda_{12}$ 
is smaller than intracomponent couplings 
 in $\lambda_{11}$ and $\lambda_{22}$
(the $U(2)$ symmetric case corresponds to 
$\lambda_{11}=\lambda_{22}=\lambda_{12}  = \lambda$). 

At the classical level, this symmetry is spontaneously 
broken in the vortex core 
and two type-I NG modes are trapped in the vortex. 
At the quantum level, the $U(1)^2$ symmetry 
is recovered and 
there are two 
Tomonaga-Luttinger liquids
as far as the so-called Tomonaga-Luttinger parameter in the isospin sector
is greater than 1 (See Appendix \ref{App:Bethe}). 

It is straightforward to extend our model to 
an ($N+1$)-component BEC with $U(N) \times U(1)$ symmetry.
There appear NG modes $S^{N-1}\simeq S^1 \ltimes 
{\mathbb C}P^{N-1}$, where $S^1$ and ${\mathbb C}P^{N-1}$ are 
type-I and II NG modes at the classical level. 
At the quantum level, the $U(1)$ symmetry is recovered 
with the Tomonaga-Luttinger liquid, 
while the $SU(N)$ symmetry remains broken 
and ${\mathbb C}P^{N-1}$ modes remain 
as type-II NG modes, as 
ensured by the Eisenberg-Lieb theorem 
\cite{PhysRevLett.89.220403}.  

Finally, fate of  type-II NG modes  
localized on domain walls 
\cite{Kobayashi:2014xua,Takahashi:2014vua}
and skyrmion lines 
\cite{Kobayashi:2014eqa} in the presence of 
quantum effects 
is one interesting future problem to explore.

{\bf Acknowledgements}

The work of MN is supported in part by Grant-in-Aid for Scientific Research (No. 25400268) 
and by the ``Topological Quantum Phenomena'' 
Grant-in-Aid for Scientific Research 
on Innovative Areas (No. 25103720)  
from the Ministry of Education, Culture, Sports, Science and Technology 
(MEXT) of Japan. 
SU is supported by the Swiss National Science Foundation under MaNEP and division II.
The work of WV has been supported by a Global Engagement for Global Impact programme funded by EPSRC.


\newpage

\appendix

\section{Derivation of the effective action}\label{app:eff-ac}

\begin{figure}[h]
\centering
\includegraphics[width=0.9\columnwidth]{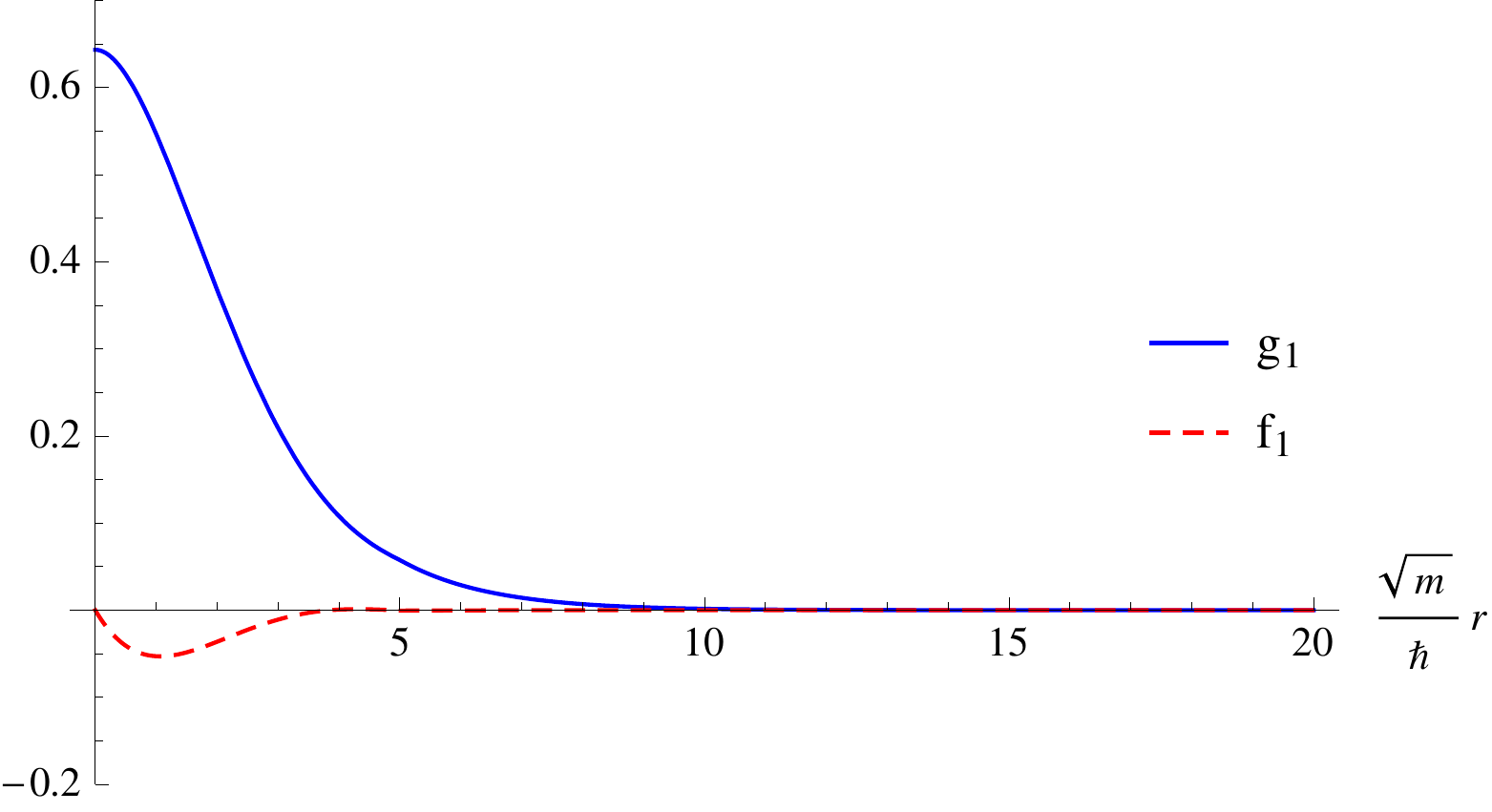} 
\caption{Numerical solution for the first order corrections to the profile functions $f_0$ and $g_0$. $\mu=2$, $\mu_0 = \lambda = \lambda_0=3$, $\kappa = 2.3$, $m = m_0$.}%
\label{fig:firstorderprofiles}%
\end{figure}

Assuming the following derivative expansion for the profile functions as explained in the text
\begin{eqnarray}
 &&g = g_0+ i\hbar    \eta^\dagger \dot \eta \, g_1+\mathcal O(\partial_t^2,\partial_z^2), \nonumber  \\
  &&f = f_0+  i\hbar    \eta^\dagger \dot \eta \,f_1+\mathcal O(\partial_t^2,\partial_z^2). \nonumber
\end{eqnarray}
we obtain the following equations:
\begin{eqnarray}
 0&=& 
-\frac{\hbar^2}{2m_0}\left[ \frac{1}{r^2}f_1 - \triangle f_1   \right] 
 +  \mu_0 f_1  - 3 \lambda_0 f_0^2f_1 \non  
&& -  \kappa  f_1 g_0^2- 2 \kappa  f_0 g_0 g_1  \nonumber \\
 0&=&
    g_0  -\frac{\hbar^2}{2m}\left[ -\triangle g_1 \right] 
+ \mu g_1  - 3 \lambda g_0^2g_1 \non
&&  -  \kappa  g_1 f_0^2 -  2\kappa  g_0 f _0 f_1 .
\label{eq:firstorder}
\end{eqnarray}
Since the non-homogeneous term in the effective action is exponentially small, the functions $f_1$ and $g_1$ are also exponentially small. A numerical solution for these quantities is shown in Fig.~\ref{fig:firstorderprofiles}. Using Eq.~(\ref{eq:coeffs}), we can evaluate numerically the coefficients $\alpha$ and $K$. Using the values of the parameters as indicated in the figures captions, we have, for example:
\begin{equation}
\alpha = 0.684, \quad \beta = 1.09,
\end{equation}
where we have taken $\hbar = m = 1$.

\section{Derivation of Bethe ansatz equations}\label{App:Bethe}
Here we sketch a derivation of the Bethe ansatz equations
in two-component bosons, which can be solved by the so-called
nested Bethe ansatz method.
Our derivation is based on one in Ref. 
\cite{deguchi2000thermodynamics}, which  discusses the Fermionic Hubbard model.

We first consider $N$ particle wavefunction as
\beq
&&\Psi(x_1,\cdots,x_N;\sigma_1,\cdots,\sigma_N)=\nonumber\\
&&\sum_{\{P\}}
A_{\sigma_{Q_1}\cdots\sigma_{Q_N}}(k_{p_1},\cdots,k_{p_N})
e^{i\sum_{j=1}^Nk_{p_j}x_{Q_j}},\\
&&1\le x_{Q_1}\le x_{Q_2}\cdots \le x_{Q_N}\le L,
\eeq
where $\{P\}$ denotes possible permutations.
From the continuity of the wavefunction and its first derivative
at equal positions as $x_1=x_2$,
we have
\beq
&&A_{\sigma_{Q_1}\cdots,\sigma_{Q_i},\sigma_{Q_{i+1}},\cdots\sigma_{Q_N}}(
k_{p_1},\cdots,k_{p_{i+1}},k_{p_i},\cdots,k_{p_N})\nonumber\\
&&=\frac{-i\lambda}{k_{p_{i}}-k_{p_{i+1}}+i\lambda}\times\nonumber\\
&&A_{\sigma_{Q_1}\cdots,\sigma_{Q_i},\sigma_{Q_{i+1}},\cdots\sigma_{Q_N}}(
k_{p_1},\cdots,k_{p_{i}},k_{p_{i+1}},\cdots,k_{p_N})\nonumber\\
&&+\frac{k_{p_{i}}-k_{p_{i+1}}}{k_{p_{i}}-k_{p_{i+1}}+i\lambda}\times\nonumber\\
&&A_{\sigma_{Q_1}\cdots,\sigma_{Q_{i+1}},\sigma_{Q_{i}},\cdots\sigma_{Q_N}}(
k_{p_1},\cdots,k_{p_{i}},k_{p_{i+1}},\cdots,k_{p_N}).\nonumber\\
\label{eq:a3}
\eeq
We next introduce
\beq
|k_{p_1},\cdots,k_{p_N}\rangle=
\sum_{\{\sigma_i\}=\uparrow,\downarrow}
A_{\sigma_{Q_1},\cdots\sigma_{Q_N}}(
k_{p_1},\cdots,k_{p_N})\nonumber\\
\times|\sigma_1,\cdots,\sigma_N\rangle,
\label{eq:ket}
\eeq
and 
\beq
&&Y^{(a,b)}(x)=\frac{-i\lambda}{x+i\lambda}I+\frac{x}{x+i\lambda}\Pi^{(a,b)},\\
&&|\sigma_1,\cdots,\sigma_{b},\sigma_a,\cdots,\sigma_N\rangle,
=\Pi^{(a,b)}|\sigma_1,\cdots,\sigma_{a},\sigma_b,\cdots,\sigma_N\rangle,
\nonumber\\
\eeq
with the identify operator $I$.
Then, Eq. \eqref{eq:a3} can be expressed as
\beq
Y^{(a,b)}(k_{p_i}-k_{p_{i+1}})
|k_{p_1},\cdots,k_{p_i},k_{p_{i+1}}\cdots,k_{p_N}\rangle=\nonumber\\
|k_{p_1},\cdots,k_{p_{i+1}},k_{p_{i}}\cdots,k_{p_N}\rangle.
\eeq
We next consider the periodic boundary condition:
$\Psi(x_1,\cdots,x_{i-1},1,x_{i+1},\cdots,x_N)
=\Psi(x_1,\cdots,x_{i-1},L+1,x_{i+1},\cdots,x_N)$.
Then, 
\beq
A_{\sigma_{Q_1}\cdots\sigma_{Q_N}}(k_{p_1},\cdots,k_{p_N})
=\nonumber\\
e^{ip_1L}
A_{\sigma_{Q_2}\cdots\sigma_{Q_N},\sigma_{Q_1}}(k_{p_2},\cdots,k_{p_N},
k_{p_1}).
\eeq
By substituting the above into Eq. \eqref{eq:ket}, we have
\beq
&&|k_{p_1},\cdots,k_{p_N}\rangle=\nonumber\\
&&e^{ip_1L}\prod_{j=0}^{N-2}
X^{(1,N-j)}(k_{p_1}-k_{p_{N-j}})
|k_{p_1},\cdots,k_{p_N}\rangle,\\
&&X^{(j,k)}(x)=\Pi^{(j,k)}Y^{(j,k)}(x).
\eeq 
By introducing
\beq
&&S\equiv\prod_{j=1}^{N-1}X^{(a,N-J)}(k_{p_1}-k_{p_{N-j}})\nonumber\\
&&=-X^{(a,N)}(k_{p_1}-k_{p_{N}})\cdots X^{(a,2)}(k_{p_1}-k_{p_2})
\Pi^{(a,1)}, \non
\eeq
which satisfies 
\beq
\text{Tr}_a S=-\prod_{j=0}^{N-2}X^{(1,N-j)}(k_{p_1}-k_{p_{N-j}}),
\eeq
we have
\beq
|k_{p_1}\cdots k_{p_N}\rangle=-e^{ik_{p_1}L}\text{Tr}_{a} S
|k_{p_1}\cdots k_{p_N}\rangle.
\label{eq:bethe-ansatz}
\eeq
By using Pauli matrices, we have
\beq
X^{(a,j)}(l-l_j)=\frac{1}{l-l_j+i\lambda}[
-i\lambda\Pi^{(a,j)}+(l-l_j)
I^{a}\otimes
I^{(j)}]\nonumber\\
=\frac{1}{l-l_j+i\lambda}\times\nonumber\\
\begin{pmatrix} (l-l_j-i\lambda/2)I_j-i\lambda\sigma^z_j/2 
& i\lambda\sigma^-_j \\
 i\lambda\sigma^+_j &   (l-l_j-i\lambda/2)I_j+i\lambda\sigma^z_j/2
\end{pmatrix},
\nonumber\\
\eeq
with $l_j=k_{p_j}-i\lambda/2$ since
\beq
\Pi^{(a,j)}=\frac{1}{2}
\begin{pmatrix}
I_j+\sigma^z_j/2 & 2\sigma^-_j \\
 2\sigma^+_j &   I_j-\sigma^z_j/2
\end{pmatrix}.
\eeq
We next introduce the so-called transfer matrix
\beq
T^{(a)}(l)&=&X^{(a,N)}(l-l_N)\cdots
X^{(a,1)}(l-l_1
)\nonumber\\
&\equiv&
\begin{pmatrix}
A(l)& B(l)\\
C(l)&D(l)
\end{pmatrix}.
\eeq
By considering a ``vacuum'' state as 
$|\text{vac}\rangle=|1,\cdots,1\rangle$,
we obtain
\beq
&&A(l)|\text{vac}\rangle=\prod_{j=1}^N
\left(\frac{l-l_j-i\lambda}{l-l_j+i\lambda}\right)
|\text{vac}\rangle,\\
&&C(l)|\text{vac}\rangle=0,\\
&&D(l)|\text{vac}\rangle=\prod_{j=1}^N\left(
\frac{l-l_j}{l-l_j+i\lambda}\right)|\text{vac}\rangle.
\eeq
Since $Y^{(a,b)}$ and $T^{(a)}$ satisfy the celebrated Yang-Baxter relation
\beq
Y^{(a,b)}(l-\mu)T^{(a)}(l)T^{(b)}(\mu)
=T^{(a)}(\mu)T^{(b)}(l)Y^{(a,b)}(l-\mu),\nonumber\\
\eeq
it is straightforward to show the following commutation relations:
\beq
&&B(l)B(\mu)=B(\mu)B(l),
\label{eq:com1}\\
&&A(\mu)B(l)=\frac{l-\mu-i\lambda}{l-\mu}B(l)A(\mu)
+\frac{i\lambda}{l-\mu}B(\mu)A(l),\nonumber\\ 
\label{eq:com2}\\
&&D(\mu)B(l)=\frac{\mu-l-i\lambda}{\mu-l}B(l)D(\mu)
+\frac{i\lambda}{\mu-l}B(\mu)D(l).\nonumber\\
\label{eq:com3}
\eeq

We finally put the following ansatz for the wavefunction:
\beq
|k_{p_1}\cdots k_{p_N}\rangle=B(l_1)\cdots B(l_M)|\text{vac}
\rangle.
\eeq
By substituting the above into Eq. \eqref{eq:bethe-ansatz}
and 
using the commutation relations \eqref{eq:com1},
\eqref{eq:com2}, and \eqref{eq:com3}, 
we obtain the following Bethe ansatz equations:
\beq
e^{ik_{j}L}=-\prod_{l=1}^{N}\frac{k_j-k_l+i\lambda}{k_j-k_l-i\lambda}
\prod_{\beta=1}^{M}\frac{k_j-l_{\beta}-i\lambda/2}{k_j-l_{\beta}+i\lambda/2},
\label{eq:bethe-eq1}\\
\prod_{j=1}^{N}\frac{l_{\alpha}-k_j-i\lambda/2}{l_{\alpha}-k_j+i\lambda/2}
=-\prod_{\beta=1}^{M}
\frac{l_{\alpha}-l_{\beta}-i\lambda}{l_{\alpha}-l_{\beta}+i\lambda}.
\eeq
Finally, by taking the logarithm of the above equations,
we obtain
Eqs.
\eqref{eq:bae1} and \eqref{eq:bae2}.

\section{Derivation of dressed energies}
For general $N$ and $M$,
the ground state energy can be expressed as
\beq
E/L&=&\frac{1}{L}\left(\sum_{j=1}^Nk_j^2-\mu N-hM+\frac{hN}{2}\right)
\nonumber\\
&=&\int_{-\Lambda}^{\Lambda}dk(k^2-\mu+h/2)\rho(k)
-h\int_{-\Omega}^{\Omega}dl\sigma(l),
\eeq
where $h$ is a magnetic field, and
$\sigma(l)$ is the density of the isospin rapidities.
From the Bethe ansatz equations, we can show 
\beq
\vec{\rho}(k,l)=\vec{\rho}_0(k,l)+K(k,l|k',l')\otimes\vec{\rho}(k',l')
\eeq
where 
\beq
&&\vec{\rho}(k,l)=(\rho(k),\sigma(l))^T,\\
&&\vec{\rho}(k,l)=(1/(2\pi),0)^T,\\
&&K(k,l:k'l')=\frac{1}{\pi}
\begin{pmatrix}
\frac{\lambda}{\lambda^2+(k-k')^2} & 
\frac{-\lambda/2}{(\lambda/2)^2+(k-l')^2} \\
  \frac{\lambda/2}{(\lambda/2)^2+(l-k')^2} &  
\frac{-\lambda}{\lambda^2+(l-l')^2} 
\end{pmatrix},
\eeq
and $\otimes$ means the integral over the same variables.

We next introduce
\beq
\vec{\varepsilon}_0(k,l)=(k^2-\mu-h/2,0)^T.
\eeq
Then,
the ground state energy can be rewritten as
\beq
E/L&=&\vec{\varepsilon}_0^T\otimes\vec{\rho},\nonumber\\
&=&\vec{\rho}_0^T\otimes\vec{\varepsilon},
\eeq
where $\vec{\varepsilon}=(\epsilon_c(k),\epsilon_s(l))^T$
is the dressed energy.
This satisfies
\beq
\vec{\varepsilon}=\vec{\varepsilon}_0+K^T\otimes \vec{\varepsilon},
\eeq
which is nothing but Eqs. \eqref{eq:ds-charge} and \eqref{eq:ds-spin}.

\section{Tomonaga-Luttinger liquid in two-component Bose gases}
Here, we consider the following Hamiltonian in two-component
bosons,
\beq
H=\int dx \sum_{\sigma}\left(\frac{\hbar^2}{2m}\partial
\Psi^{\dagger}_{\sigma}\partial\Psi_{\sigma}
-\mu\Psi^{\dagger}_{\sigma}\Psi_{\sigma}
+\frac{\lambda}{2}(\Psi^{\dagger}_{\sigma}\Psi_{\sigma})^2\right)
\nonumber\\+\kappa \Psi^{\dagger}_{1}\Psi_{1}
\Psi^{\dagger}_{2}\Psi_{2}.
\eeq
Here, we are interested in the miscible case, that is,
\beq
\lambda>\kappa ,\\
\lambda>0.
\eeq
In what follow, we consider the situation that
two components have the same density,
$\rho_{1}=\rho_{2}$.
Then, we can apply the bosonization to this ``bosonic'' system. 
In the bosonization for bosons \cite{giamarchi2003quantum},
we can use the following bosonization 
\beq
\Psi^{\dagger}_{\sigma}=\left(\rho_0-\frac{1}{\pi}\nabla\phi_{\sigma}\right)^{1/2}\left[e^{-i\theta_{\sigma}}
+\sum_{p=\pm1}e^{2ip(\pi\rho_0x-\phi_{\sigma})-i\theta_{\sigma}}
\right],\\
\rho_{\sigma}=\Psi^{\dagger}_{\sigma}\Psi_{\sigma}=
\left(\rho_0-\frac{1}{\pi}\nabla\phi_{\sigma}\right)
+\rho_0\sum_{p=\pm1}e^{2ip(\pi\rho_0x-\phi_{\sigma})},
\eeq
where $\phi_{\sigma}$ and $\theta_{\sigma}$
are conjugate fields in the bosonization.

By applying the above dictionary to the our Hamiltonian,
we have
\beq
H\to
 \frac{1}{2\pi}\int dx\sum_{\mu=\rho,\sigma}
\left[u_{\mu}K_{\mu}(\nabla\theta_{\mu})^2
+\frac{u_{\mu}}{K_{\mu}}(\nabla\phi_{\mu})^2
\right]\nonumber\\
+2\kappa \rho_0^2\int dx\cos2\sqrt{2}\phi_{\sigma},
\eeq
where we introduce
\beq
\phi_{\rho}=\frac{1}{\sqrt{2}}(\phi_{1}+\phi_{2}),\\
\phi_{\sigma}=\frac{1}{\sqrt{2}}(\phi_{1}-\phi_{2}),
\eeq
and the similar relations for $\theta$ fields.
Here, $u_{\mu}$ is the velocity and $K_{\mu}$ is the so-called
Luttinger parameter,
and they are given by
\beq
K_{\rho}=\frac{K}{\sqrt{1+\kappa K/\pi u}},\\
K_{\sigma}=\frac{K}{\sqrt{1-\kappa K/\pi u}},\\
u_{\rho}=u\sqrt{1+\kappa K/\pi u},\\
u_{\rho}=u\sqrt{1-\kappa K/\pi u},\\
K=\sqrt{\frac{\pi^2\rho_0}{\lambda M}},\\
u=\sqrt{\frac{\lambda\rho_0}{M}}.
\eeq

As can be seen from the above bosonization Hamiltonian, 
if there is no coupling between up and down particles 
($\kappa =0$),
the Hamiltonian is quadratic, which is called
Tomonaga-Luttinger Hamiltonian.

In the presence of $\kappa $,
we also have $\cos\phi_{\sigma}$, and therefore
physics becomes non-trivial.
To see what happens in the presence of this coupling,
we can consider the renormalization group analysis, which
is one of the most powerful approaches in one dimensional world.
By applying the perturbative renormalization group in the above
Hamiltonian, we have the following renormalization group equations
\cite{giamarchi2003quantum},
\beq
\frac{dK_{\sigma}(l)}{dl}=-\frac{\lambda^2_{12}(l)
K^2_{\sigma}(l)}{2},\\
\frac{d\kappa (l)}{dl}=(2-2K_{\sigma}(l))
\kappa (l).
\eeq
We note that the above are essentially equivalent to renormalization 
group equations
in the so-called BKT transition.
We can easily check that $\cos\phi_{\sigma}$ is locked since 
$\kappa $ flows to strong coupling if $K_{\sigma}<1$
while $\cos\phi_{\sigma}$ is irrelevant if  $K_{\sigma}>1$.

In a weakly-interacting bosonic system,
we naturally expect that  $K_{\sigma}>1$,
and therefore the fixed point is the Tomonaga-Luttinger liquid phase.
Namely, even if we consider the quantum fluctuations,
we still expect that the low-energy excitations in the system are linear
gapless.

\bibliography{reference}

\end{document}